\documentclass[twocolumn,showpacs,aps,pra]{revtex4}
\usepackage{epsfig,amssymb}

\begin{document}

\title{Alkali adsorbate polarization on conducting and insulating surfaces probed with Bose-Einstein condensates}
\author{J.~M. McGuirk, D.~M. Harber, J.~M. Obrecht, and E.~A. Cornell$^\ast$}
\affiliation{JILA, National Institute of Standards and Technology and Department of Physics, \\
University of Colorado, Boulder, Colorado 80309-0440}
\date{\today}

\begin{abstract}
A magnetically trapped $^{87}$Rb Bose-Einstein condensate is used as a sensitive probe of short-range electrical forces.  In particular, the electric polarization of, and the subsequent electric field generated by, $^{87}$Rb adsorbates on conducting and insulating surfaces is measured by characterizing perturbations to the magnetic trapping potential using high-Q condensate excitations.  The nature of the alterations to the electrical properties of Rb adsorbates is studied on titanium (metal) and silicon (semiconductor) surfaces, which exhibit nearly identical properties, and on glass (insulator), which displays a smaller transitory electrical effect.  The limits of this technique in detecting electrical fields and ramifications for measurements of short-range forces near surfaces are discussed.
\end{abstract}

\maketitle

With the recent advances in trapping technology for Bose-condensed neutral atoms, both in extremely stable free-space magnetic traps \cite{harber2002} and in stable surface microtraps \cite{chips,vuletic2003}, ultracold atoms near macroscopic surfaces have become appealing systems for studies of quantum information, high sensitivity interferometry, and precision force measurements in the range of $1-100\,\mu$m.  Ultracold atom-based measurements of Casimir-Polder forces \cite{hinds1993,aspect1996,shimizu2001,vuletic2003}, searches for new physics at small distance scales \cite{dimop2003}, chip-based atom interferometry and quantum computing \cite{chips}, and some studies of lower dimensional Bose-Einstein condensates \cite{grimm2003} all rely on working with atoms in close proximity to macroscopic surfaces.  One of the properties that makes isolated neutral atoms appealing for these types of experiments is their relative insensitivity to electrical perturbations.  While ground state alkali atoms have small DC polarizabilities and electric interactions are usually extremely small, electric dipole forces are not always negligible.

In close proximity to nonuniform charge distributions, small electric fields can have large gradients.  When working with ultracold atoms near surfaces, it is likely that some of the atoms will become stuck to the surface.  Depending on the substrate it is possible for the electrons in the adatoms and substrate to redistribute themselves, leading to nonzero electric fields from initially neutral objects.  Using a $^{87}$Rb Bose-Einstein condensate, we report the quantitative measurements of the electric fields from polarized $^{87}$Rb adsorbates on silicon, titanium, and BK7 glass surfaces in the range of $5-30\,\mu$m from the surfaces.  These fields pose a potentially serious impediment to measurements relying on ultracold atoms near surfaces, leading to spurious forces, decoherence, and heating.  However, this work also demonstrates a sensitive technique for probing surface-based electric fields and presents the possibility of using these perturbations to manipulate ultracold atoms in novel ways.  This paper is organized in the following manner:  we briefly describe the sticking of atoms to substrates, show how condensates can be sensitive probes of surface-induced perturbations and describe quantitatively how these perturbations influence condensate behavior, present quantitative measurements of Rb dipoles on three different surfaces, and finally discuss ramifications of and applications for these results.

\begin{figure}
\leavevmode
\epsfxsize=3.375in
\epsffile{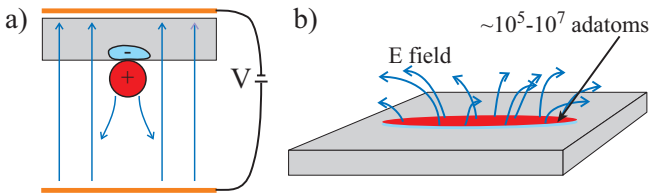}
\caption{\label{fig:cartoon} (color online). a) Schematic depiction of an adsorbed Rb atom on a conducting substrate (not to scale).  The adatom in part relinquishes its valence electron to the substrate, leaving a positively charged ion bound to the substrate with a negative image charge in the substrate.  The typical adatom-image charge separation is $\sim 5$~\AA.  Electric field plates are added to the setup to enhance the effect of ionized adsorbates.  The applied electric field may be oriented in either direction to add constructively or destructively to the surface dipolar field.  b) Roughly elliptical footprint of adsorbates following deposition of a number of elongated condensates, producing large electric field gradients near the surface.}
\end{figure}

The electronic structure of an atom changes as it sticks to a surface.  For a ground state alkali atom, the lowest lying \emph{S} and \emph{P} levels of the valence electron will interact with the energy bands of the substrate, producing new hybridized energy levels for the bonded atom \cite{scheffler2000}.  If part of the renormalized atomic levels falls below the Fermi energy of the substrate, then the valence electron resides partially in the substrate as well as the atom.  The net effect of the orbital hybridization is a fractional charge transfer to the substrate.  The resultant atom-substrate bond is somewhat arbitrarily labelled, \emph{e.g.} ``ionic'' or ``covalent,'' depending on the relative electronegativities of the bond constituents \cite{bonds}.  For the purposes of this work, we merely note that if there is a charge transfer from adatom to substrate then the resultant bond has at least some ionic character.  Substrates in which atom-substrate bonds have significant ionic character are roughly those with work functions comparable to or greater than the ionization energy of the adatom.  For $^{87}$Rb this is 4.2~eV, as compared to the work functions of Si (4.8~eV) and  Ti (4.3~eV) \cite{workfn}.  Rb is expected to be electropositive on Si and Ti surfaces at room temperature, but not on an insulator such as glass (where there is little fractional charge transfer and the atom-substrate bond is primarily due to van der Waals forces).  The effect of the fractional charge transfer to the substrate essentially is to produce a dipole comprised of a positively charged ion with a negative image charge inside the substrate (schematically shown in Fig.~\ref{fig:cartoon}).  The fractional charge transfer for Rb on Si or Ti is expected to be somewhat less than unity, with a typical effective distance (bond length) of $\sim 5$~\AA\ between the positive ion and the negative image charge \cite{scheffler2000}.  Although there is expected to be no significant charge transfer between Rb and glass, the electron orbitals of Rb adsorbates on glass are nevertheless perturbed, thus altering the polarizability and possibly inducing a small dipole moment.

If a magnetically trapped atom is brought near a partially ionized adatom, the electric field produced by the surface dipole will polarize the trapped atom, and it will experience an attractive potential according to $U_{\mbox{\scriptsize{dip}}}(r) = - \frac{\alpha}{2} E_{\mbox{\scriptsize{dip}}}(r)^2$, where $\alpha$ is the ground state DC polarizability of the atom (79~mHz/(V/cm)$^2$ \cite{polar}) and $E_{\mbox{\scriptsize{dip}}}(r)$ is the dipolar field at distance $r$ from the surface.  For certain numbers and distributions of ionized adatoms, this potential can become large in comparison to the Casimir-Polder force and even to the magnetic confining forces.  With the partial pressures of Rb present in our vacuum chamber, it would take several years for a significant amount of Rb to build up, and even then a spatially uniform distribution of adsorbates is unlikely to produce a large enough concentration of dipoles to create significant field gradients.  However, if a number of condensates are stuck to the surface, either purposefully or accidentally during the process of performing a surface-based measurement, then a nonuniform areal concentration can be achieved with a strong spatial dependence (Fig.~\ref{fig:cartoon}b).  Finally, if the surface coverage becomes large enough so that the adsorbate distribution becomes more homogeneous, the force should diminish as the electric field becomes smaller and more uniform.

\begin{figure}
\leavevmode
\epsfxsize=3.375in
\epsffile{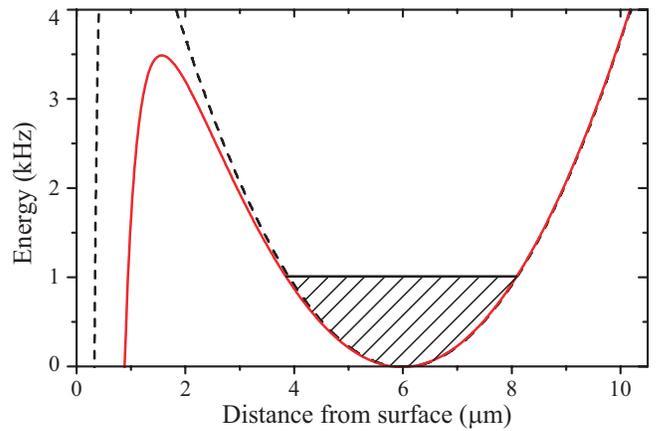}
\caption{\label{fig:potential} (color online). Typical potential experienced by a condensate magnetically trapped near a surface.  The dotted line is the sum of the harmonic magnetic trapping potential and the Casimir-Polder potential, and the solid curve includes the calculated electric potential from $\sim 10^7$ Rb atoms adsorbed on a conducting surface in a pattern $\sim 4 \mu$m~$\times\ 150 \mu$m.  The hatched region is the region occupied by a typical condensate at equilibrium.}
\end{figure}

When a condensate in a harmonic magnetic confining potential is brought near the surface, the harmonic potential is perturbed by any surface-related forces.  Condensates in harmonic magnetic traps are excellent tools for measuring small perturbations to the local potential.  Condensates in harmonic magnetic traps (\emph{e.g.} Ioffe-Pritchard type traps) with weak spring constants and low anharmonicity can support several high-Q collective mechanical excitations.  In particular, the transverse dipole mode (center-of-mass radial slosh) decouples from internal degrees of freedom in a harmonic trap, leading to minimal damping \cite{stringari1999}.  Additionally, the transverse monopole mode (breathing mode) is also undamped in an axisymmetric harmonic trap \cite{dalibard2002}. With excitation lifetimes of several seconds, the quality factor Q of these modes can be as high as $10^4$, which allows the trap frequency to be determined with extremely high precision.  Quantitative measurements of surface-based perturbing forces are made by measuring the induced trap frequency shifts.  If the curvature of the perturbing potential is negative, then the trap frequency is lowered, and if the perturbation's curvature is positive the frequency increases.  (Negative and positive curvatures typically correspond to attractive and repulsive forces respectively near a surface.)  This method is similar to the operation of atomic force microscopes for the measurement of small forces \cite{afm}.  The transverse dipole mode is discussed exclusively in this work, but we have observed reasonable quantitative agreement between measurements using both dipole and radial monopole modes.  Moreover the damping of these modes is related to the harmonicity of the trap, and as perturbations make the trap more anharmonic, condensate excitations are damped more rapidly, providing a possible second measure of perturbations to the harmonic potential.  Fig.~\ref{fig:potential} shows a typical potential experienced by atoms trapped near a surface.

The apparatus used to measure surface-related forces is described in detail in Ref.~\cite{harber2003}.  The experiment consists of a highly elongated, axially symmetric condensate created in a conventional magnetic trap (frequencies of $5.6 \times 216 \times 216$~Hz, giving an $\sim39:1$ aspect ratio) whose long axis is parallel to the surfaces to be studied and perpendicular to gravity.  Nearly pure condensates containing $\sim 10^5$ atoms are created far from the surfaces ($\sim 600\,\mu$m) and are then smoothly brought near the surface by means of an applied magnetic field normal to the surface.  (The applied magnetic field, when summed with the linear magnetic radial gradient, acts to shift the center of the trap in the direction of the applied field.)  Once the condensate is at the desired distance from the surface ($5-30\,\mu$m), the radial dipole mode normal to the surface is excited by applying a brief, nonadiabatic magnetic field gradient, \emph{i.e.} by displacing the trap center by $\sim 4\,\mu$m for 0.5~ms.  The amplitude of the resulting sloshing motion is $\sim 2\,\mu$m.  Destructive images are taken at various times in the oscillation cycle, and sinusoidal fits are applied to the position of the condensate in order to extract the trap frequency.  Far from the surface, where the potential is mostly harmonic, typical damping times of the dipole mode are several seconds.  In order to characterize different substrates, the coils of the magnetic trap are physically shifted to perform experiments over a different surface.

To test that perturbations to the trapping potential are induced by electric dipoles bound to the surface \cite{magnetic}, a uniform external electric field, $E_{\mbox{\scriptsize{app}}}$, is applied normal to the surface (see Fig.~\ref{fig:cartoon}a).  This field is generated by placing large copper plates above and below the glass vacuum cell and applying up to $\pm 150$~V to the lower capacitor plate while grounding the upper.  In this way, the total electric potential becomes $U_{\mbox{\scriptsize{el}}}(r) = - \frac{\alpha}{2} |{\bf E}_{\mbox{\scriptsize{app}}} + {\bf E}_{\mbox{\scriptsize{surf}}}({\bf r})|^2$.  If the applied field is significantly larger than the surface field, the effective potential is $U_{\mbox{\scriptsize{el}}}(r) \simeq - \alpha {\bf E}_{\mbox{\scriptsize{app}}} \cdot {\bf E}_{\mbox{\scriptsize{surf}}}({\bf r})$.  The net effect of the applied field is to amplify any surface-related fields and also to change the spatial scaling of the resultant potential.  For instance, if $E_{\mbox{\scriptsize{surf}}}(r) \sim 1/r^3$, then $U_{\mbox{\scriptsize{el}}}(r) \sim 1/r^6$ with no applied field and $\sim 1/r^3$ with a strong external field.

We measure the electric field of adsorbates by depositing a number of atoms on the surface and measuring the trap frequency as a function of voltage applied to the capacitor plates.  In order to ensure that the surface is initially free from adsorbates, the atom-surface distance calibration (see Ref.~\cite{harber2003}) is performed over a different location on the surface by shifting the condensate by $25\,\mu$m parallel to the surface using a transverse magnetic field.  Deposition of atoms is accomplished simply by moving the trap center into the substrate with an applied magnetic field.

\begin{figure}
\leavevmode
\epsfxsize=3.375in
\epsffile{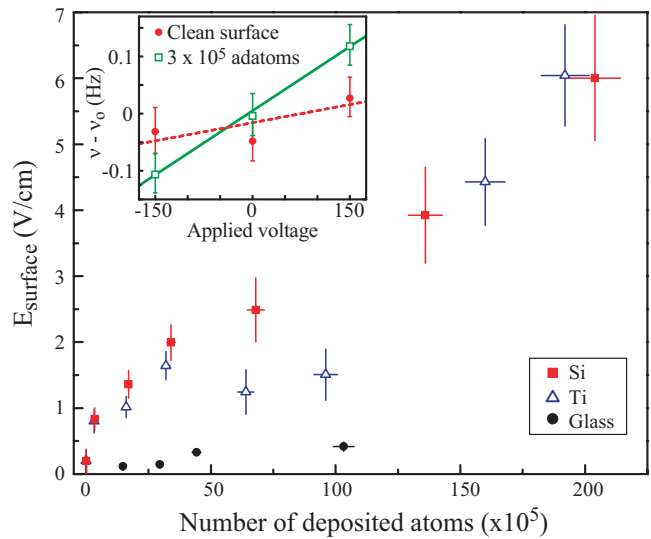}
\caption{\label{fig:N} (color online). Surface-based electric field as a function of atoms stuck to the surface.  Si ({\tiny$\blacksquare$}) and Ti ($\triangle$) substrates exhibit similar behavior, roughly linearly increasing with adsorbate number, while the glass substrate ({\large$\bullet$}) shows only a small effect of adsorbates.   The distance between the center of the magnetic trap and the surface is fixed to $10\,\mu$m.  Vertical error bars denote statistical errors only and do not represent systematic uncertainties, notably uncertainties in $E_{\mbox{\scriptsize{app}}}$ and in the power law of $E_{\mbox{\scriptsize{surf}}}$.  The inset shows a typical plot of frequency versus applied voltage before ({\large$\bullet$}) and after ({\tiny$\square$}) one condensate has been deposited on Ti ($\nu_\circ = 216.5$~Hz).  There is a measurable effect from the atoms of only one condensate adsorbing.  The lines are weighted fits to extract the frequency shift per applied voltage.}
\end{figure}

The inset of Fig.~\ref{fig:N} shows the trap frequency as a function of applied voltage before and after one condensate containing $\sim 3 \times 10^5$ atoms has been stuck to the Ti surface.  This procedure is repeated with an increasing number of condensates stuck to the three different substrates, and the results are characterized by studying the normalized change in trap frequency, $\delta \nu / \nu_\circ$.  Using the analysis techniques described below, we extract the value of the surface-based electric field from the slope of the frequency shift versus applied voltage measurements.  This field is shown in Fig.~\ref{fig:N} as a function of number of atoms stuck to the surface.  As expected from their similar work functions, Si and Ti show nearly identical behavior; any discrepancy between the two may be due to small differences in their work functions, which lead to different electric dipole moments for the partially ionized Rb adsorbates, or by the exact spatial arrangement of the adatoms on each substrate.  Even for the largest numbers of adsorbed atoms, the surface coverage is at most only a few percent of a monolayer.

The goal of the data analysis is to extract both the distance dependence of the adsorbate electric field as well as the magnitude of the electric field from a known number of adatoms.  In order to extract quantitative results from trap frequency shifts, we make use of two approximate methods: an approximation for a classical, point-like oscillator, and a one-dimensional Gross-Pitaevskii equation simulation.  The point-like approximation is made by linearly expanding the perturbing potential about the center of the trap (center of the oscillation) and replacing the spring constant with an effective spring constant (see Ref.~\cite{gady1996}).  The result is that the fractional shift in trap frequency $\nu$ is 
\begin{equation}
\frac{\delta\nu}{\nu_\circ} \simeq - \frac{1}{2k_\circ} \frac{\partial^2 U_{\mbox{\scriptsize{el}}}(r)}{\partial r^2},
\end{equation}
where the spring constant $k_\circ = (2 \pi \nu_\circ)^2m$ is $2.7 \times 10^{-19}$~kg/s$^2$.  The advantage of this approximation is manifest: a readily calculable relation that predicts both amplitude and spatial dependence of the electric fields.  However, this method fails to account for the macroscopic size of a condensate, and the approximation begins to break down when the amplitude of the oscillation is not significantly less than the distance to the surface.  The expansion gives improved results if it is made about the ``distance of closest approach'' of the condensate, \emph{i.e.} the center of the condensate at the inner turning point of the oscillation, as this point represents the largest perturbation experienced by the condensate.  Thus, for a polarized Rb atom attached to the surface, the electric field falls off as $1/r^3$ and the fractional frequency shift $\delta \nu/\nu_\circ$ falls off as $\sim 1/r^8$ with no applied field and $\sim 1/r^5$ with a uniform external field.  If more adsorbates accrue in a spatially homogeneous manner, then the power of the distance dependence decreases.  The point-like approximation provides a nice heuristic relation, but if exact results are desired a better calculation is required.  A full calculation propagating the condensate wavepacket with the Gross-Pitaevskii equation yields a more accurate answer, but is computer intensive and does not offer an intuitive result.  We use this method to check the results given by the point-like method.  In the end, both methods fail to give exact results without knowing the full two-dimensional spatial distribution of adatoms, although both methods give reasonable results that are consistent with each other.

\begin{figure}
\leavevmode
\epsfxsize=3.375in
\epsffile{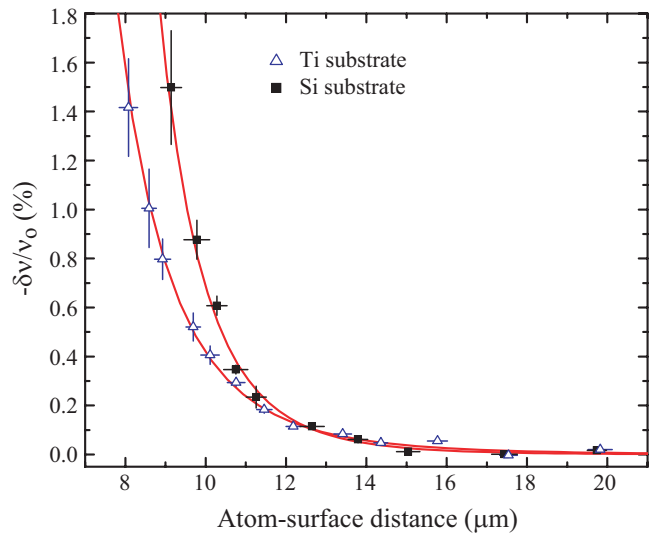}
\caption{\label{fig:distance} (color online).  Frequency shift as a function distance between the magnetic trap center and the surface, after a large number of adsorbates ($> 2 \times 10^7$) had accumulated on the surface.  Si ({\tiny$\blacksquare$}) and Ti ($\triangle$) substrates exhibit similar behavior.  Solid lines are fits giving powers of $1/r^{2.3}$ for the electric field near Si and $1/r^{2.0}$ for Ti, according to the point-like approximate frequency shift method (see text).  These powers are consistent with a highly elongated, quasi-one dimensional distribution of dipoles.}
\end{figure}

\begin{figure}
\leavevmode
\epsfxsize=3.375in
\epsffile{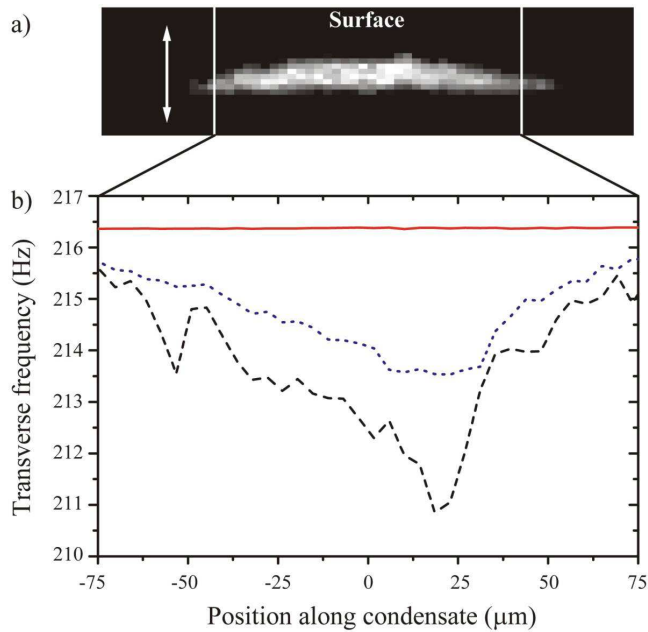}
\caption{\label{fig:spatial} (color online). a)  Absorption image of a condensate undergoing dipolar oscillations near Si.  The curvature along the axial direction (long axis) is due to an inhomogeneous distribution of adatoms, which results in different radial trap frequencies along the axial direction.  For imaging, the condensate is expanded radially by a factor of about five.  b) Axial (parallel to the surface) spatial profile of trap frequency for a condensate near a Si substrate with no applied field after a large number of atoms have adhered to the surface.  The three lines represent magnetic trap centers 19.7~$\mu$m (solid), 9.8~$\mu$m (dotted), and 9.1~$\mu$m (dashed) from the surface.  The sharp dip in frequency is indicative of a greater number of adatoms at that spatial location on the surface.  One would expect a distribution of adatoms that mirrors the elliptical distribution of the deposited condensates, with an axial Thomas-Fermi radius of $\sim 150\,\mu$m.  However, the stronger attractive surface potential near the center of the condensate might further attract atoms to stick there, thus altering the expected distribution.  The increased noise in the data is due to fitting rows of single pixels in an image rather than averaging over an entire condensate.  This technique treats the transverse oscillation of each axial segment of the elongated condensate as independent from the adjacent segment, which involves the implicit assumption that the ``stiffness'' of the condensate does not impede the oscillations.}
\end{figure}

It is impossible to obtain any value for the dipole moment of a single Rb adatom without knowing the distance dependence of the trap frequency shifts (and thus the electric field).  In order to obtain the distance dependence, a study of frequency shift as a function of distance from the surface was performed in the absence of applied electric fields.  This experiment was performed after a large number of atoms where stuck to the surface so that a small number of additional atoms adsorbing during the course of the measurement would not significantly alter the potentials.  The results of this study are shown in Fig.~\ref{fig:distance}.  Fitting the power law of frequency shift using the above point-like approximation, gives the result that the electric field falls off as $1/r^{2.3}$ for Ti and $1/r^{2.0}$ for Si, indicating that the spatial distribution of adatoms is between a point distribution ($1/r^3$) and a line of dipoles ($1/r^2$).  This result is consistent with a two-dimensional adatom distribution, which is demonstrated further by a transverse spatial analysis of the trap frequency (Fig.~\ref{fig:spatial}).

By fitting each column of pixels across the cloud in a condensate image, the trap frequency can be determined at each spatial location along the long axis of the condensate, provided the interrogation time for the measurement is shorter than the axial trapping period to prevent motional averaging.  In this manner, the surface potential is probed along the length of the condensate.  Fig.~\ref{fig:spatial} shows a small region with a more pronounced perturbation of the trap frequency, implying that there is a greater concentration of adsorbates at that location.  One would expect an approximately Thomas-Fermi shaped distribution (inverted parabola), but because atoms are attracted more strongly to regions with larger numbers of adsorbates, inhomogeneities perhaps can build up in a run-away adsorption process.

From the distance dependence of the trap frequency (Fig.~\ref{fig:distance}), one can also extract the magnitude of a single adatom-image charge dipole.  Obtaining a precise numerical value is complicated by a number of factors.  First, while the point-like approximation is surprisingly good even at small atom-surface separations, it is not accurate enough to extract precise quantitative information, especially considering the spatial extent of the condensate.  Second, in the regime where the distance measurements were performed, there was a large number of atoms stuck to the surface, and the uncertainty in adsorbate number is large.  Also, the electric field depends on the distribution of adsorbates on the surface.  Since this is evidently a nontrivial distribution (Fig.~\ref{fig:spatial}), the electric field will have different near- and  far-field spatial dependence.  However, one can use the spatial behavior from the distance measurements and extract rough quantitative values from the measurements with applied electric field, as these measurements have less dependence on atom distribution.  This method requires knowledge of the exact electric field produced by applying a certain voltage to the capacitor plates, which is complicated by the glass vacuum cell, the titanium substrate holder, and the substrates themselves.  Using a commercial field calculation program (Ansoft Maxwell SV), we have calculated the electric field near the surfaces within 10\%, which is approximately the same level as the error in atom number.  These errors are dwarfed by the uncertainty from the spatial distribution of adatoms.  With these considerations, we obtain a value of $\sim 1 \times 10^{-29}$~C\,m (3~debye) for the dipole moment of a single Rb adatom on a Si or Ti substrate.  Within our error bars, we cannot differentiate Si and Ti.  A single atom's dipole moment produces an electric field of only $1\,\mu$V/cm at a distance of $10\,\mu$m and a gradient of 3~mV/cm$^2$.  We have observed an electric field, produced by a distribution of a large number of adatoms, as large as 10~V/cm at a distance of $10\,\mu$m.

On the other hand, Rb adsorbates on glass behave significantly differently.  Trap frequency measurements near glass show only a small effect from adsorbed Rb atoms (see Fig.~\ref{fig:N}), implying that, as expected, Rb does not form ionic bonds with glass.  The electric fields observed may be attributed to the fact that Rb-glass bonds still do perturb the electronic state of the Rb adatoms, leading to a slight electric dipole moment.  It is difficult to characterize accurately the magnitude of the effect on glass.  This is due in part to the extremely small dipole moment of Rb on glass.  Additionally, van der Waals bonds such as those binding Rb to glass typically have binding energies less than 0.5~eV, and some of the Rb may desorb or diffuse across the surface during the course of a measurement.  The best measurements we make place the dipole moment of Rb on glass to be five to ten times less than Rb on Ti or Si.

When presented with such small dipole moments, the immediate question is: what are the limits of this technique in detecting surface-fixed impurities?  Clearly, it is far easier to detect isolated charges on insulating surfaces, where there is no image charge present.  For instance, the field produced by a single electron is 140~mV/cm at $10\,\mu$m.  This field is large enough to produce a $10^{-3}$ shift in trap frequency at a distance of $3\,\mu$m from the surface, and detection of this field can easily be enhanced with a externally applied uniform field.  A single ionized adatom on a conducting surface, however, produces a significantly smaller electric field that falls off more rapidly with distance.  The dipole moment of a single adatom is difficult to detect, but could be seen with an applied electric field of 1 kV/cm at a distance of $\leq 4\,\mu$m.  (These results assume a reasonable sensitivity limit of $\delta \nu / \nu_\circ \geq 10^{-5}$.)  Clearly, this method is sufficient to detect single charged particles and polarized objects attached to substrates and could be a useful detector of such.  Additionally, this method should be sensitive to other surface related electrical phenomena such as patch effects (patches with small differences in work function) \cite{patch}, dopant inhomogeneities in semiconductors, or even inhomogeneities in the dielectric constant of insulators like glass.

Electric fields from polarized adsorbates pose a potentially serious systematic for experiments operating near surfaces, especially those bent on measuring substrate-related forces with high precision.  As noted above, the strength of the dipolar forces observed in this work near conducting and semiconducting surfaces have been more than three orders of magnitude greater than the Casimir-Polder force.  It would seem that a precise Casimir-Polder measurement between alkalis and conducting or semiconducting surfaces is unlikely to be successful, even if care were taken to avoid deposition of atoms on the surface.  It is nearly impossible to avoid some adsorption during the course of performing a measurement and the accompanying calibrations, and only a small number of adatoms ($< 10^4$) is sufficient to pollute a measurement.  The same reasoning applies to other substrate-related measurements, including measurements searching for spin-gravity couplings, short-range deviations from $1/r^2$ gravity, or equivalence principle violations \cite{dimop2003}.  The limits that could be placed on these forces would be significantly less stringent in the presence of ionized adsorbates, lessening the contributions to limits on new physics from these experiments.  Additionally, an inhomogeneous potential from ionized adatoms can be harmful to experiments aimed at precision interferometry or quantum computing using microchip traps \cite{chips}, or possibly even adversely affect experiments with trapped ions \cite{wineland2000}.  Inhomogeneities from adsorbates also can lead to decoherence, excite unwanted condensate modes that can lead to heating, and cause condensate fragmentation (a possible explanation of effects seen in Ref.~\cite{grimm2003}).  Finally, frequency shifts on the order of $\sim 10^{-18}$ in atomic fountain clocks could possibly be produced by electric fields generated by alkali adsorbates on copper microwave cavities, for certain cavity geometries \cite{gibble}.

In principle, it is possible to remove the adatoms, but in practice it is nontrivial.  In an ultrahigh vacuum (UHV) environment, the atoms will desorb eventually given time.  One would expect desorption to be expedited by heating the substrate or by illuminating the substrate with ultraviolet (UV) light.  However, the temperature necessary to be effective is generally not conducive to the UHV environment required to produce a condensate, and most common vacuum cell materials are opaque to UV light.  We have measured the desorption time for Si at room temperature illuminated with a 150~W visible halogen bulb through the pyrex vacuum cell (which absorbs the UV spectral components below $\sim 340$~nm), and found the time constant to be large, in excess of three days.  Even if a bright blue or UV laser were used every experimental cycle to clean the surface, it is unlikely the surface could be kept sufficiently clean for the required sensitivity to surface forces.  In fact, applying a halogen UV light source (100~mW/cm$^2$ peaked at 365~nm) to remove adatoms from glass appears to have made the electrical forces much \emph{worse}, either by ionizing the adsorbates or ionizing other surface contaminants in the process of desorbing atoms.  Laser ablation would probably clean the surface, but this method is also not conducive to required vacuum or to careful surface calibration.  It appears that the most likely configuration for performing short-range force measurements with condensates is to use insulating substrates such as glass, or if bulk material properties are to be studied, then coating the substrate with a thin layer of insulating material to prevent adatom charge transfer would be effective.  Because there is still a small effect from Rb adatoms on glass, care needs to be taken to minimize the number of atoms that stick to the substrate.  We have observed that, contrary to its behavior on Ti and Si substrates, Rb desorbs from glass with a time constant of $\sim 12-24$~hours at room temperature, and this could possibly be accelerated by operating at only a slightly higher temperature.  Given these caveats, it should be possible to minimize any electrical effects to at least an order of magnitude less than the Casimir-Polder force, permitting a sensitive measurement of the force over the length scale of $3-10\,\mu$m.

Electric fields produced by polarized adatoms might prove to be useful tools for atomic manipulation.  The ability to control the deposition of atoms as well as control the strength of atom-adatom interactions leads one to consider the prospect of using patterned adsorbate structures to manipulate condensates in various ways.  For instance, atoms deposited in a small patch on a conducting surface can be used, with an applied electric field, to form a repulsive barrier in the center of the magnetic trap, thus creating a double-well potential whose barrier height is rapidly adjustable by means of the applied electric field.  One could use such a repulsive barrier for a number of experiments including Josephson oscillations, quantum information studies, and as a switchable beamsplitter in a magnetic waveguide interferometer.  An example of a more exotic structure that could be created is to create an atom ``grating'' on the surface.  This could be accomplished by interfering two blue-detuned (repulsive) lasers on the surface and then depositing atoms onto the surface.  Atoms will be repelled from intensity maxima and will deposit preferentially at the minima, thereby creating a grating structure from which a condensate can be diffracted.  Electrostatic forces have already been used in a microchip trap \cite{schmied2003}, and one could extend this principle to make shallow two dimensional electrostatic traps relying on surface adsorbates.

In conclusion, we have identified and systematically measured the effect of the fractional charge transfer from alkali atoms adsorbing on conducting and semiconducting substrates.  At short ranges, the electric field gradients from the dipoles formed by partially ionized adatoms and image charges in the substrate are large and can alter magnetic trapping potentials significantly.  This effect poses a serious obstacle to many condensate-based experiments near surfaces, especially those requiring sensitive force measurements.  On the other hand, this work demonstrates that neutral condensates can be sensitive electric field probes, and furthermore partial ionization of adsorbed atoms potentially could be leveraged to create novel structures for the manipulation of ultracold atoms. 

This work was supported by the NSF and NIST and is based upon work supported under a NSF Graduate Research Fellowship.


\begin{thebibliography}{Hi!}
\bibitem{bonds}
According to the common categorization, Rb-Si bonds are barely categorized as ionic, and Rb-Ti is called polar covalent.  For the purposes of this work, one can simply think of Rb-Si bonds being slightly more ``ionic in character'' than Rb-Ti, leading to a slightly larger fractional charge transfer from the adatom.
\bibitem{workfn}
The work function of glass is not well-defined, but the bandgap between the valence and conduction bands is much larger than the ionization energy of the adatoms.
\bibitem{magnetic}
We have also seen several areas on the surface that contribute magnetic irregularities to the potential from contaminants either in or on the substrate.  Magnetic forces are distinguished from electric forces through application of the external electric field, which modifies perturbations from electric surface effects but not from magnetic contaminants.  Magnetic contaminants are probably not intrinsic and can be avoided by using more rigorous surface preparation techniques.
\end{thebibliography}
\end{document}